\def\mycmd{2}

\if\mycmd1
\documentclass[11pt,onecolumn, draftcls]{IEEEtran}
\else 
\documentclass[lettersize,journal]{IEEEtran}
\fi
\usepackage[utf8]{inputenc}
\usepackage{tabularx}
\usepackage{fontawesome}
\usepackage{threeparttable}

\usepackage{colortbl}    
\usepackage{pifont}     

\usepackage{algorithm}
\usepackage{algpseudocode}

\usepackage[british]{babel}
\usepackage{kotex}
\usepackage{multicol}
\usepackage{array,ragged2e}
\usepackage{float}
\usepackage{mathtools}
\usepackage{amsmath}
\usepackage{amsthm}
\usepackage{amssymb}
\usepackage{graphicx}
\usepackage{wasysym}
\usepackage{setspace}
\usepackage{color}
\usepackage{bm}
\usepackage{cases}
\usepackage{makecell}
\usepackage{tikz}
\usepackage{flowchart}
\usepackage{amsfonts}
\usepackage{steinmetz}
\usetikzlibrary{matrix,shapes,arrows,positioning,chains}
\usepackage{lmodern,babel,adjustbox,booktabs,multirow}
\usepackage{makecell}
\usepackage{pbox}
\usepackage{epsfig}
\usepackage{tcolorbox}
\usepackage{enumitem}

\allowdisplaybreaks[4]

\makeatletter
\newcommand{\multiline}[1]{%
  \begin{tabularx}{\dimexpr\linewidth-\ALG@thistlm}[t]{@{}X@{}}
    #1
  \end{tabularx}
}

\floatstyle{ruled}
\newfloat{algorithm}{tbp}{loa}
\providecommand{\algorithmname}{Algorithm}
\floatname{algorithm}{\protect\algorithmname}

\theoremstyle{plain}

\theoremstyle{plain}

\theoremstyle{plain}

\usepackage{epsfig}
\usepackage[caption=false,font=normalsize,labelfont=sf,textfont=sf]{subfig}
\usepackage{cite}
\usepackage{stfloats}
\usepackage{graphicx}
\usepackage{multirow}
\usepackage{array}
\usepackage{enumerate}
\usepackage{graphicx}
\usepackage{times}
\usepackage{algorithm}
\usepackage{algpseudocode}
\theoremstyle{remark}

\makeatother

\algrenewcommand\algorithmicindent{1.0em}%
\providecommand{\lemmaname}{Lemma}
\providecommand{\propositionname}{Proposition}

\providecommand{\theoremname}{Theorem}
\providecommand{\theoremname}{Definition}
\newcommand{\rom}[1]{\uppercase\expandafter{\romannumeral #1\relax}}

\newcounter{problem}
\newcounter{save@equation}
\newcounter{save@problem}

\definecolor{lightergray}{gray}{0.9}
\definecolor{ForestGreen}{RGB}{34,139,34}  


\numberwithin{save@problem}{subsection}
\numberwithin{save@equation}{subsection}

\begin{document}
\title{Group-wise Semantic Splitting Multiple Access for Multi-User Semantic Communication}
\author{Jungyeon Koh,~\IEEEmembership{Member,~IEEE}, Hyeonho Noh,~\IEEEmembership{Member,~IEEE}, and Hyun Jong Yang,~\IEEEmembership{Senior Member,~IEEE}
\thanks{Jungyeon Koh is with the Department of Electrical Engineering, Pohang University of Science and Technology (POSTECH), Korea (e-mail: jungyeon.koh@postech.ac.kr). Hyeonho Noh is with the Department of Information and Communication Engineering, Hanbat National University, Republic of Korea (e-mail: hhnoh@hanbat.ac.kr).
Hyun Jong Yang is with the Department of Electrical and Computer Engineering, Seoul National University, Korea (email: hjyang@snu.ac.kr). Jungyeon Koh and Hyeonho Noh contributed equally to this work.
}
}

\maketitle
\begin{abstract}\label{abstract}
In this letter, we propose a group-wise semantic splitting multiple access framework for multi-user semantic communication in downlink scenarios. The framework begins by applying a balanced clustering mechanism that groups users based on the similarity of their semantic characteristics, enabling the extraction of group-level common features and user-specific private features. The base station then transmits the common features via multicast and the private features via unicast, effectively leveraging both shared and user-dependent semantic information. To further enhance semantic separability and reconstruction fidelity, we design a composite loss function that integrates a reconstruction loss with a repulsion loss, improving both the accuracy of semantic recovery and the distinctiveness of common embeddings in the latent space. Simulation results demonstrate that the proposed method achieves up to $3.26\times$ performance improvement over conventional schemes across various channel conditions, validating its robustness and semantic efficiency for next-generation wireless networks.
\end{abstract}



\section{Introduction}
\label{sec:introduction}
Next-generation wireless networks, including beyond 5G and 6G, are expected to connect massive numbers of devices and deliver vast volumes of data. This explosive growth in demand exacerbates the longstanding issue of spectrum scarcity, as the available frequency bands remain fundamentally limited. 
To address this limitation, semantic communication~\cite{chaccour2024less, shao2024theory} has emerged as a promising paradigm. Instead of transmitting all raw bits, semantic communication extracts and transmits compact semantic representations, such as high-level features or embeddings, to substantially reduce the communication load while preserving task-relevant performance.

Building on this foundation, recent research has investigated how to further enhance semantic communication performance using conventional wireless techniques. In particular, multiple access methods---such as orthogonal multiple access (OMA) \cite{Ding24_TWC, Liu24_TCCN, Noh24_wiopt} and rate-splitting multiple access (RSMA) \cite{cheng2023resource, xu2025rate, liu2025generative, lu2024utility, dizdar2024rate}---are increasingly being integrated with semantic communication. By leveraging separation in the power, frequency, or time domains, these approaches improve communication performance in multi-user scenarios.
However, these approaches overlook the heterogeneity of user data when assigning time, frequency, or power resources. Since users generate data sporadically with diverse characteristics, the grouping strategy should explicitly account for such heterogeneity. In addition, despite semantic communication’s reliance on deep learning models, existing studies have not sufficiently formulated loss functions that align with the specific multiple access strategies used during semantic training.



\begin{figure}[t]
    \centering
    \includegraphics[width=\columnwidth]{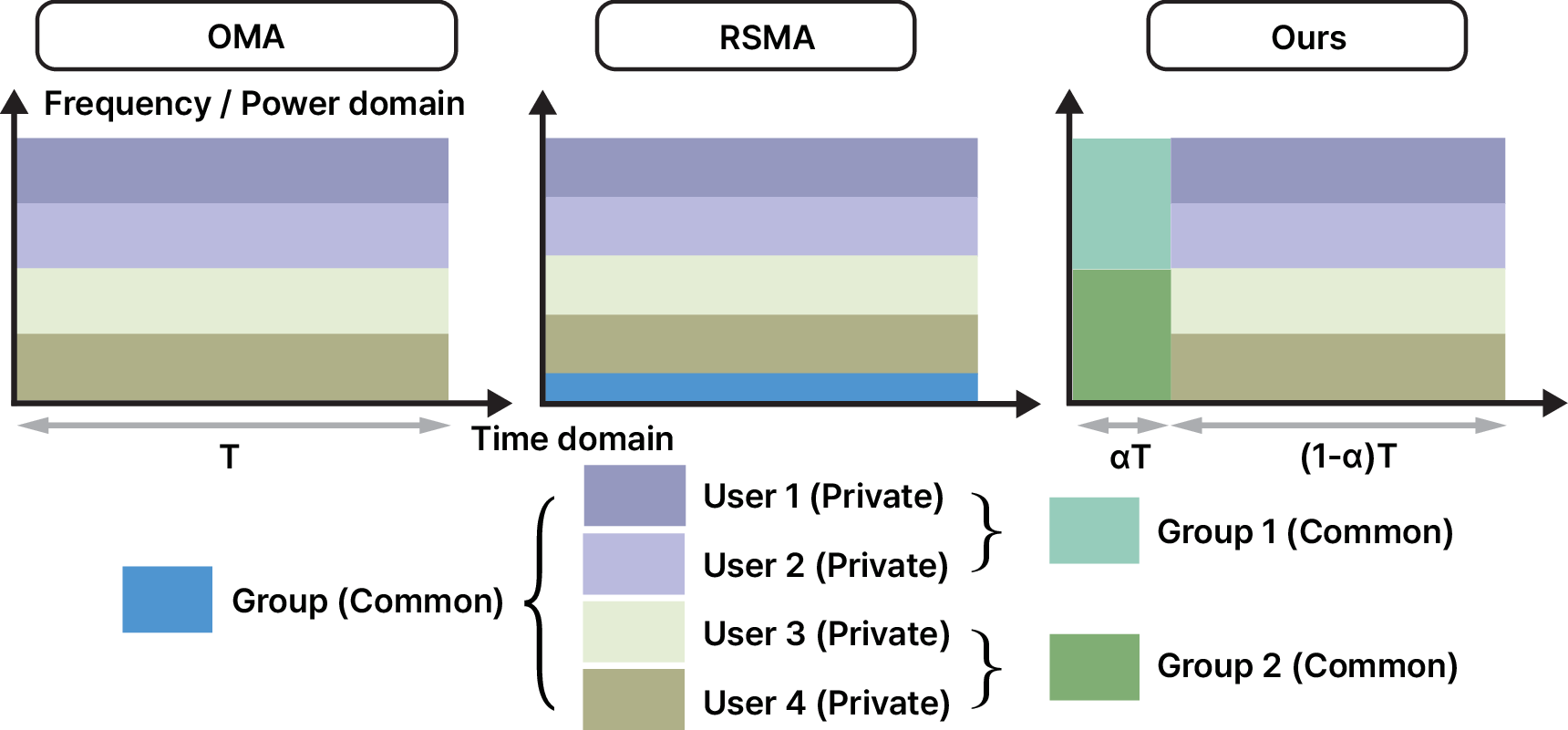}
    \caption{Comparison between conventional works and the proposed one.}
    \label{fig:comparison}
\end{figure}

In this paper, we propose a group-wise semantic split multiple access framework for multi-user downlink data transmission. 
The core idea is to group users based on the correlation in their data, and subsequently, for each group, to transmit distinct sets of common and private semantic features, as illustrated in Fig. \ref{fig:comparison}. By organizing users with correlated data into coherent groups and assigning each group its own common and private semantic features, the proposed framework enables a structured separation of representations that enhances both semantic efficiency and multi-user transmission performance.

To realize this idea, we first introduce a heuristic balanced clustering method that groups users according to the similarity of their data characteristics. This allows the transmitter to construct group-wise common features independently, ensuring that the semantic model learns more homogeneous common representations within each group.
To optimize the model for both accurate reconstruction and effective semantic separation, we design a composite loss function that integrates a reconstruction loss with a repulsion loss. The reconstruction loss guarantees that the combined common and private features can faithfully recover the original input, while the repulsion loss enhances the distinctiveness of common features by maximizing their angular and Euclidean distances in the latent space.
Simulation results demonstrate that the proposed approach achieves up to $3.26\times$ performance improvement over conventional methods, particularly in terms of reconstruction accuracy and semantic feature disentanglement.

\section{System Model}
\begin{figure*}[t]
    \centering
    \includegraphics[width=\textwidth]{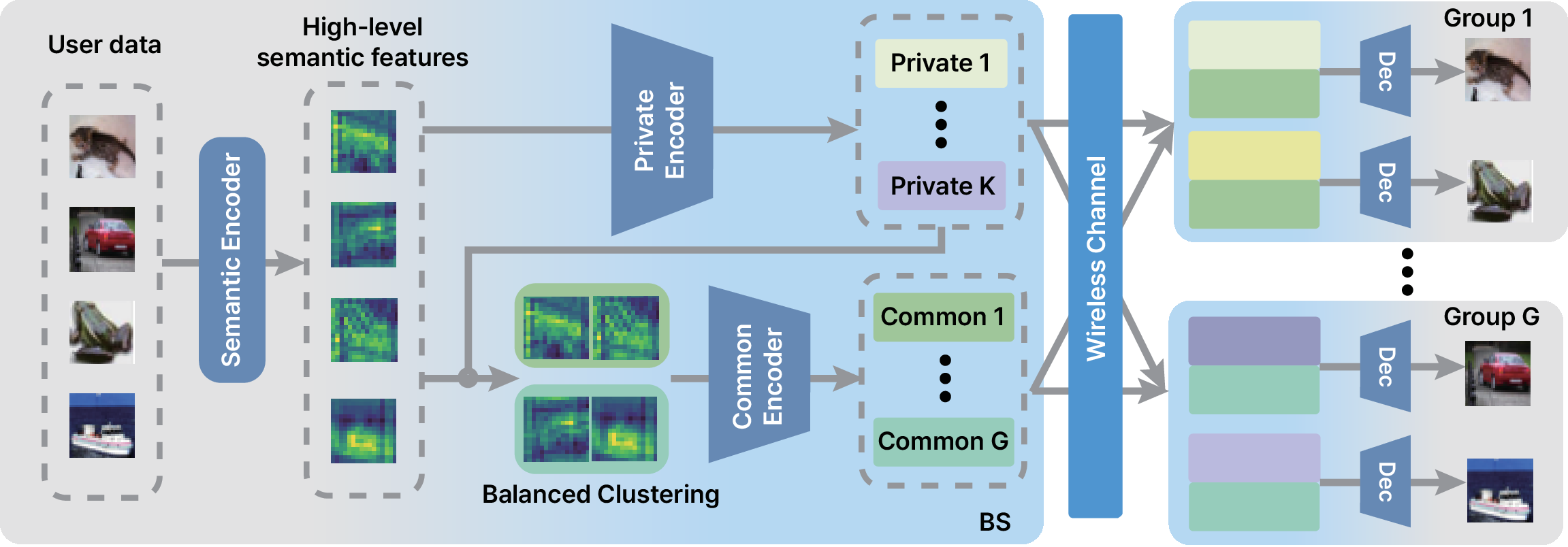}
    \caption{Overview of the proposed model.}
    \label{fig:model_architecture}
\end{figure*}


We consider a multi-user downlink semantic communication system consisting of a base station (BS) and $K$ single-antenna communication users. The BS transmits independent data to users, so that each user can perform its own task. 

\subsection{Semantic Transmitter}
As shown in Fig.~\ref{fig:model_architecture}, we denote the source data of the $k$-th user as $s_k$, which carries task-relevant semantic information.
Each source data is first processed by a semantic encoder to extract high-level semantic features, which are represented by
\begin{align}
    z_k = S(s_k; \beta),
\end{align}
where $z_k$ is the resulting semantic representation and $S(\cdot; \beta)$ is the semantic encoder with learnable parameters $\beta$. 
These extracted features are then processed to generate two components: private features, which contain user-specific semantic information, and common features, which capture common semantics across users. 
The private feature encoder extracts users' private features, which can be expressed as
\begin{align}
\mathbf{p}_k &= f_p(z_k; \theta_p),
\end{align}
where $\mathbf{p}_k \in \mathbb{R}^{L_p \times 1}$ denotes the private feature of user $k$, $L_p$ is the length of the private feature, and $f_p(\cdot; \theta_p)$ is a learnable private feature encoder parameterized by $\theta_p$.

Before extracting the common features, a clustering step is performed based on the cosine similarities of the private features $\{p_k\}_{k=1}^K$. 
Specifically, users with similar semantic characteristics are clustered into $G$ groups, denoted as $\{\mathcal{G}_1, \mathcal{G}_2, \ldots, \mathcal{G}_G\}$, where each group $\mathcal{G}_g$ contains users sharing similar private semantics. The detailed clustering algorithm is described in \ref{subsec:train_process}.

For each group $\mathcal{G}_g$, a Transformer-based encoder is employed to aggregate semantic information and extract a group-wise common feature:
\begin{align}
    \mathbf{c}_g = f_c(\{z_k\}_{k \in \mathcal{G}_g}; \theta_c), \quad g = 1, 2, \ldots, G,
\end{align}
where $\mathbf{c}_g \in \mathbb{R}^{L_c \times 1}$ denotes the common feature of group $\mathcal{G}_g$, $L_c$ is the length of the common feature, $f_c(\cdot; \theta_c)$ is the common feature encoder with the learnable parameter $\theta_c$. 
The resulting $\{\mathbf{c}_g\}_{g=1}^G$ capture group-level common semantics that benefit all users within each group and serve as communication-efficient representations shared in the system. 

\subsection{Semantic Receiver}

After obtaining the common features $\{\mathbf{c}_g\}_{g=1}^G$, the BS performs semantic feature transmission in two stages.

\textbf{Common-Feature Transmission: } In the first stage, the BS transmits the group-level common features $\{\mathbf{c}_g\}$ to users within the corresponding group $\mathcal{G}_g$. 
The received signal at user $k \in \mathcal{G}_g$ can be written as
\begin{align}
    y_k^{(c)}[l] 
    &= h_k \mathbf{c}_g[l]
    + \sum_{g' = 1, g' \ne g}^G h_k \mathbf{c}_{g'}[l]
    + n_k,
\end{align}
where $h_k$ denotes the downlink channel coefficient for user $k$, and $n_k$ represents additive white Gaussian noise (AWGN) with variance $\sigma_k^2$. 

\textbf{Private-Feature Transmission: } Next, the BS transmits user-specific private semantic features $\{\mathbf{p}_k\}$.  
The received signal at user $k$ can be represented by
\begin{align}
    y_k^{(p)}[l] 
    &= h_k \mathbf{p}_k[l]
    + \sum_{k' = 1, k' \ne k}^K h_k \mathbf{p}_{k'}[l]
    + n_k.
\end{align}

Before transmission over the wireless channel, the extracted features are power-normalized. In this work, we consider three channel models: AWGN, Rician, and Rayleigh fading.
Under the AWGN channel, the effects of fading are ignored.
For the Rayleigh fading channel, the channel coefficients follow 
$h_k \sim \mathcal{CN}(0, 1)$. 
For the Rician fading channel, the coefficients follow 
$h_k \sim \mathcal{CN}(\mu, \sigma^2)$, with $\mu = \sqrt{r/(r+1)}$ and $\sigma = \sqrt{1/(r+1)}$, 
where $r$ is the Rician factor. 

The semantic decoder aims to reconstruct \(\hat{s}_k\) based on the features received from the channel.
After receiving both the common and private semantic features, each user concatenates them and reconstructs its intended semantic information. 
The reconstructed source data of user $k \in \mathcal{G}_g$ can be expressed as 
\begin{align}
    \hat{s}_k = S^{-1}(\mathrm{CONCAT}[\mathbf{c}_g, \mathbf{p}_k]; \phi_k),
\end{align}
where $S^{-1}(\cdot; \phi_k)$ denotes the semantic decoder parameterized by $\phi_k$, which maps the concatenated semantic features back to the reconstructed source domain. 
This decoding process preserves both the common semantics within the group and the user-specific details unique to each source.

\section{Proposed Method}
We propose a composite loss function that jointly optimizes the reconstruction accuracy and latent-space separability of common features. Furthermore, we introduce a balanced clustering strategy for effective common feature grouping.

\begin{figure}[t]
    \centering
    \includegraphics[draft=false, width=0.9\columnwidth]
    {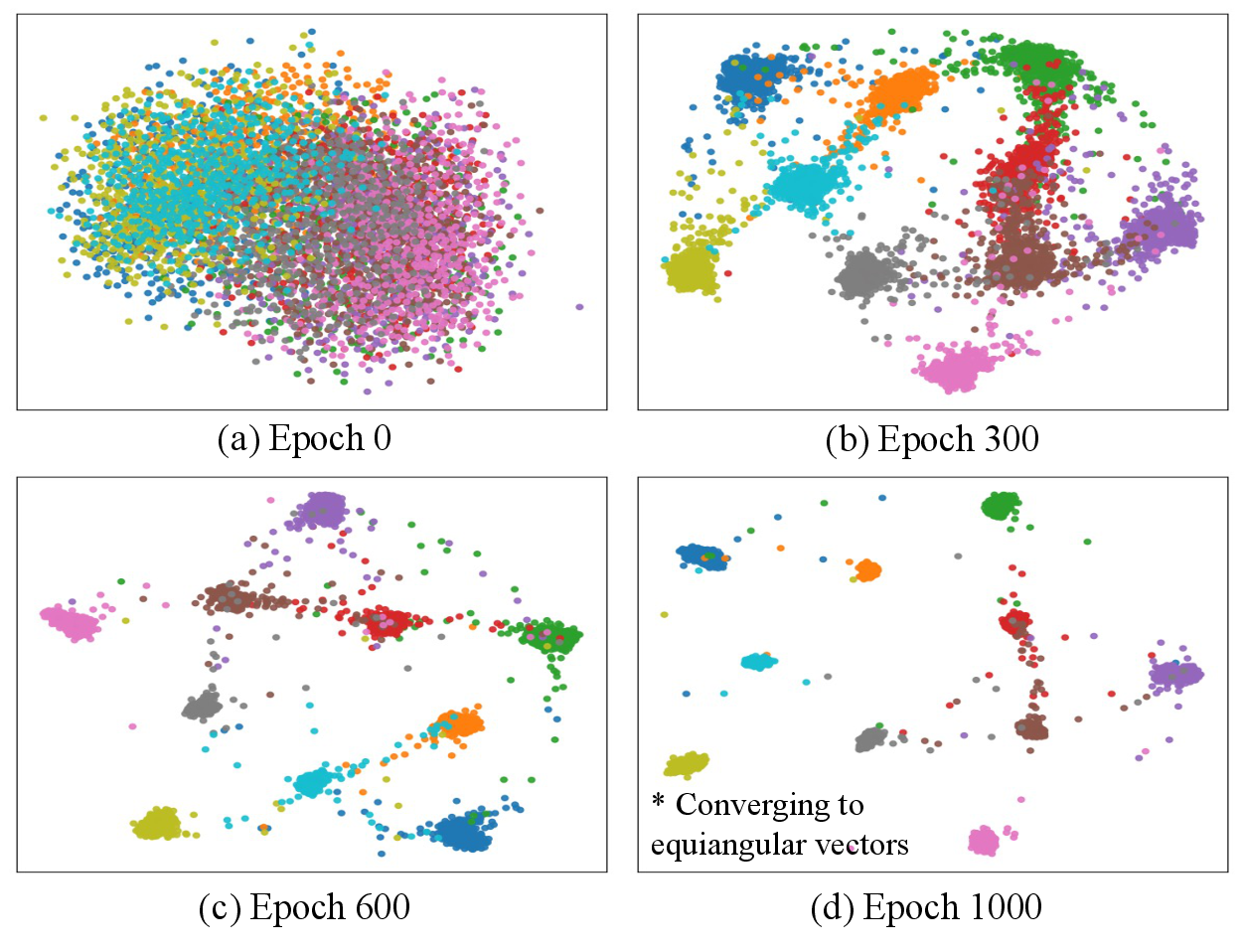}
    \caption{t-SNE plots of common features at different epochs. Notably, feature representations converge toward an equiangular, pentagon-shaped configuration.}
    \label{fig:tsne_grid}
\end{figure}

\subsection{Composite Loss Function Design} 
The proposed composite loss consists of two components: a reconstruction loss and a repulsion loss, which is written as
\begin{align}
\mathcal{L}_{\text{total}}=\mathcal{L}_{\text{recon}}+\lambda \mathcal{L}_{\text{repul}},
\end{align}
where \(\lambda\) is the weight of the repulsion term.
The reconstruction loss $\mathcal{L}_{\text{recon}}$ ensures high-quality image recovery at the receiver sides, while the repulsion loss $\mathcal{L}_{\text{repul}}$ prevents the common features---typically compressed more aggressively than private ones---from collapsing into overlapping regions in the latent space. 
Fig. \ref{fig:tsne_grid} shows a t-SNE plot of common features at different epochs. As the repulsion loss is minimized, both Euclidean and angular separations increase, resulting in more discriminative group-wise embeddings.

\textbf{Reconstruction loss}: We adopt the Charbonnier loss\cite{bruhn2005lucas} as $\mathcal{L}_{\text{recon}}$, which provides improved robustness to outliers compared with the conventional \(l_2\) loss.
\begin{align}
\mathcal{L}_{\text{recon}}=\frac{1}{K}\sum_{i=1}^K \sqrt{(s^i-\hat{s}^i)^2+\epsilon^2},
\end{align}
where \(\epsilon\) is a small constant for numerical stability.

\textbf{Repulsion loss: } The repulsion loss consists of three components: a Euclidean repulsion term, a center regularization term, and an angular repulsion term, which is expressed as
\begin{multline}
\mathcal{L}_{\text{repul}}=\underbrace{\frac{1}{G(G-1)}\sum_{\substack{i,j=1\\ i\neq j}}^{G} e^{-d_{ij}}}_{\text{Euclidean repulsion}}
+ \underbrace{\lambda_c \left\| 
\frac{1}{G} \sum_{i=1}^{G} \mathbf{g}_i 
\right\|_{2}^{2}}_{\text{Center regularization}} 
\\+ \underbrace{\frac{1}{G(G-1)}\sum_{\substack{i,j=1\\ i\neq j}}^{G} (X_{ij}-T_{ij})^{2}}_{\text{Angular repulsion}}.
\end{multline}
where \(d_{ij}=\Vert \mathbf{c}_i-\mathbf{c}_j\Vert_2^2\) denotes the Euclidean distance between common features \(\mathbf{c}_i\) and \(\mathbf{c}_j\), and \(\lambda_c\) is the weight of the center regularization term. Furthermore, \(X_{ij}=\frac{\mathbf{c}_i^\top \mathbf{c}_j}
{\Vert \mathbf{c}_i\Vert_2\,\Vert \mathbf{c}_j\Vert_2}\) represents the cosine similarity between  common features, and \(T_{ij}\) specifies the target pairwise cosine similarity among \(G\) group-level common features, which is given by 
\begin{align}
    T_{ij} =
\begin{cases}
1, & i=j, \\[2pt]
-\dfrac{1}{G-1}, & i\neq j~.
\end{cases}
\end{align}

The \textbf{Euclidean repulsion term} penalizes feature pairs that are spatially close via an exponential distance function. The \textbf{center regularization term} prevents overfitting and improves generalization to unseen data\cite{le1991eigenvalues}.
To maximize the separation between feature representations, it is important to consider not only the Euclidean distance but also the angular distance.
Motivated by findings in simplex-based subspace clustering\cite{xu2019scaled, cai2024tensorized}, 
we propose an \textbf{angular repulsion term}.
To be specific, the angular repulsion term drives the pairwise cosine similarity between any two common features toward \(-1/(G-1)\). This enforces an equiangular configuration on the hypersphere, thereby maximizing angular separability among group-wise common features. 
Since the common feature dimension $L_c$ typically satisfies \(L_c\gg G\), forming such an equiangular configuration is theoretically feasible. 

\subsection{Training Process}\label{subsec:train_process}
\textbf{Balanced clustering strategy:}
To address heterogeneous user data and improve group-wise semantic alignment, our framework adopts a label-agnostic clustering strategy. It operates solely on extracted private features, thereby avoiding any dependency on labeled information.
We first apply the K-means algorithm to obtain \(G\) initial centroids, and construct a cost matrix $\mathbf{D}$ based on cosine distances between the private features and these centroids. The resulting assignment problem is then solved using the Hungarian matching algorithm\cite{kuhn1955hungarian}, formulated as follows:
\begin{equation}
\begin{aligned}\label{eq:hungarian_alg}
\min_{\mathbf{M}} &\sum_{k=1}^{K} \sum_{g=1}^{G} d_{k,g} m_{k,g} \\
\text{s.t.} \quad &m_{k,g}\in\{0,1\}, \quad \forall k,g \\
&\sum_{g=1}^{G} m_{k,g} = 1, \quad \forall k  \\
\end{aligned}
\end{equation}
where $\mathbf{M}\in\{0,1\}^{K\times G}$ denotes the binary assignment matrix. This formulation ensures that each user feature is assigned to exactly one cluster. The complete procedure is summarized in  Algorithm~\ref{alg:balanced-clustering}.
\begin{algorithm}[t]
\caption{Balanced Clustering via K-means and Hungarian Matching}
\label{alg:balanced-clustering}
\begin{algorithmic}[1]
\Require Private Feature matrix $\mathbf{P}\in \mathbb{R}^{K \times L_p}$, number of groups $G$
\State Apply K-means with $G$ clusters to $\mathbf{P}$, obtain centroids $\mathbf{C} \in \mathbb{R}^{G \times L_p}$
\State Compute $\mathbf{D} \in \mathbb{R}^{K \times G}$
\State Solve the Hungarian matching algorithm in \eqref{eq:hungarian_alg} to obtain $\mathbf{M}\in\{0,1\}^{K\times G}$
\State Form groups: $\mathcal{G}_g \gets \{ k \mid m_{k,g} = 1 \}, \quad \forall g \in \{1, \dots, G\},\forall k \in \{1,\dots,K\}$
\Return $\mathcal{G} = \{ \mathcal{G}_1, \dots, \mathcal{G}_G \}$
\end{algorithmic}
\end{algorithm}

\textbf{Joint training algorithm:} Algorithm \ref{alg:joint-training} outlines the end-to-end training procedure of the proposed framework, which jointly optimizes both common and private feature representations.
\begin{algorithm}[t]
\caption{Joint Training Algorithm}
\label{alg:joint-training}
\begin{algorithmic}[1]
\Require Training dataset $\mathbf{S}$ and the batch size $K$
\State \textbf{Transmitter:}
\State Choose mini-batch data $\{\mathbf{s}_j\}_{j=n}^{n+K}$ from $\mathbf{S}$
\State \(z_k \gets S(s_k; \beta)\)
\State $\mathbf{p}_k \gets f_p\big(\text{FC}(\text{MaxPool}(z_k)); \theta_p\big)$
\State Derive $\mathcal{G} = \{ \mathcal{G}_1, \dots, \mathcal{G}_G \}$ using Alg.~\ref{alg:balanced-clustering}
\State $\mathbf{c}_g \gets f_c(\{z_k\}_{k \in \mathcal{G}_g}; \theta_c)$
\State Transmit $\mathbf{p}_k$ through unicast and $\mathbf{c}_g$ through multicast over the channel
\State \textbf{Receiver:}
\State $\hat{s}_k \gets S^{-1}(\mathrm{CONCAT}[\mathbf{c}_g, \mathbf{p}_k]; \phi_k)$,
\State Compute $\mathcal{L}_{\text{recon}}(s_k, \hat{s}_k)$ and 
$\mathcal{L}_{\text{repul}}(\{\mathbf{c}_g\})$,

\State Compute $\mathcal{L}_{\text{total}}=\mathcal{L}_{\text{recon}}+\lambda \mathcal{L}_{\text{repul}}$
\State Update $\{\beta, \theta_p, \theta_s, \phi_k\}$ using gradient descent with $\mathcal{L}_{\text{total}}$
\end{algorithmic}
\end{algorithm}

\section{Performance Evaluation} 
\subsection{Datasets and Simulation Environments}

\begin{table}[htbp]
\centering
\caption{Semantic Network Architecture}
\label{table:semantic_network}
\resizebox{\linewidth}{!}{
\begin{tabular}{|c|c|c|}
\hline
\textbf{} & \textbf{Layer Name} & \textbf{Units} \\

\hline
\multirow{5}{*}{\textbf{ConvNext\cite{liu2022convnet} Block($d$)}} 
& Conv2D (kernel=4, stride=2) & $128$ \\
& Conv2D (kernel=7, padding=3) & $d$ \\ 
& Conv2D (kernel=1) & $4d$ \\
& GRN (Global Response Norm) & $4d$ \\
& Conv2D (kernel=1) & $d$ \\

\hline
\multirow{2}{*}
{\textbf{Semantic Encoder}}
& $2\times$ ConvNext Block($128$) & $128$ \\
& $3\times$ ConvNext Block($256$) & $256$ \\

\hline
\multirow{2}{*}
{\textbf{Private Encoder}}
& MaxPool2D (kernel=4) & - \\
& Dense & $L_p$ \\

\hline
\multirow{4}{*}{\textbf{Common Encoder}}
& $3\times$ ConvNext Block($512$) & $512$ \\
& MaxPool2D (kernel=4) & - \\
& 2$\times$Transformer Encoder & 512 (4 heads) \\
& Dense & $L_g$ \\

\hline
\multirow{7}{*}{
\textbf{Semantic Decoder}} & Dense & $512$ \\
& Upsample (scale=4, bilinear) & - \\
& TransposeConv2D (kernel=4, stride=2) & 256 \\
& $3\times$ ConvNext Block($256$) & 256 \\
& TransposeConv2D (kernel=4, stride=2) & $128$ \\
& $3 \times$ ConvNext Block($128$) & $128$ \\
& TransposeConv2D (kernel=4, stride=2) & $3$ \\
& $2\times$ ConvNext Block($3$) & $3$ \\
\hline
\end{tabular}
}
\end{table}

The proposed method is trained and evaluated on the CIFAR10\cite{krizhevsky2009learning} dataset, and its architecture is summarized in Table~\ref{table:semantic_network}. All network modules use ReLU activation functions. Training is performed for $1000$ epochs using the Adam optimizer with a learning rate of $10^{-4}$. Each training batch contains \(50\) images, which are evenly divided into \(10\) groups of \(5\) images to simulate multi-user semantic encoding. During training, the SNR is uniformly sampled between 12 dB and 18 dB. Model performance is evaluated in terms of peak signal-to-noise ratio (PSNR) and perceptual loss, where the latter is computed using a pretrained VGG network\cite{johnson2016perceptual}.

We evaluate the proposed method against three baselines.
\begin{itemize}[leftmargin=*]
    \item \textbf{ViT-based}: a recently introduced semantic communication model based on a Vision Transformer (ViT).
    \item \textbf{Private-only}: an ablated version of our architecture without group-level common feature sharing.
    \item \textbf{JPG-LDPC}: a conventional separate source-channel coding (SSCC) framework that uses JPEG for source compression and LDPC for channel coding. 
\end{itemize}
To ensure fairness, all semantic communication methods transmit the same number of symbols per image. For the SSCC baseline, LDPC parameters and modulation formats are selected such that the total number of transmitted bits closely matches that of the proposed system for each source. Reported PSNR values for SSCC corresponds to the best performance achieved across various combinations of code rates and modulation configurations.


\subsection{Comparison with Conventional Works}
\begin{figure}[h]
    \centering 
    \includegraphics[draft=false, width=\if 1 \mycmd 0.6 \else 1.0 \fi \columnwidth]{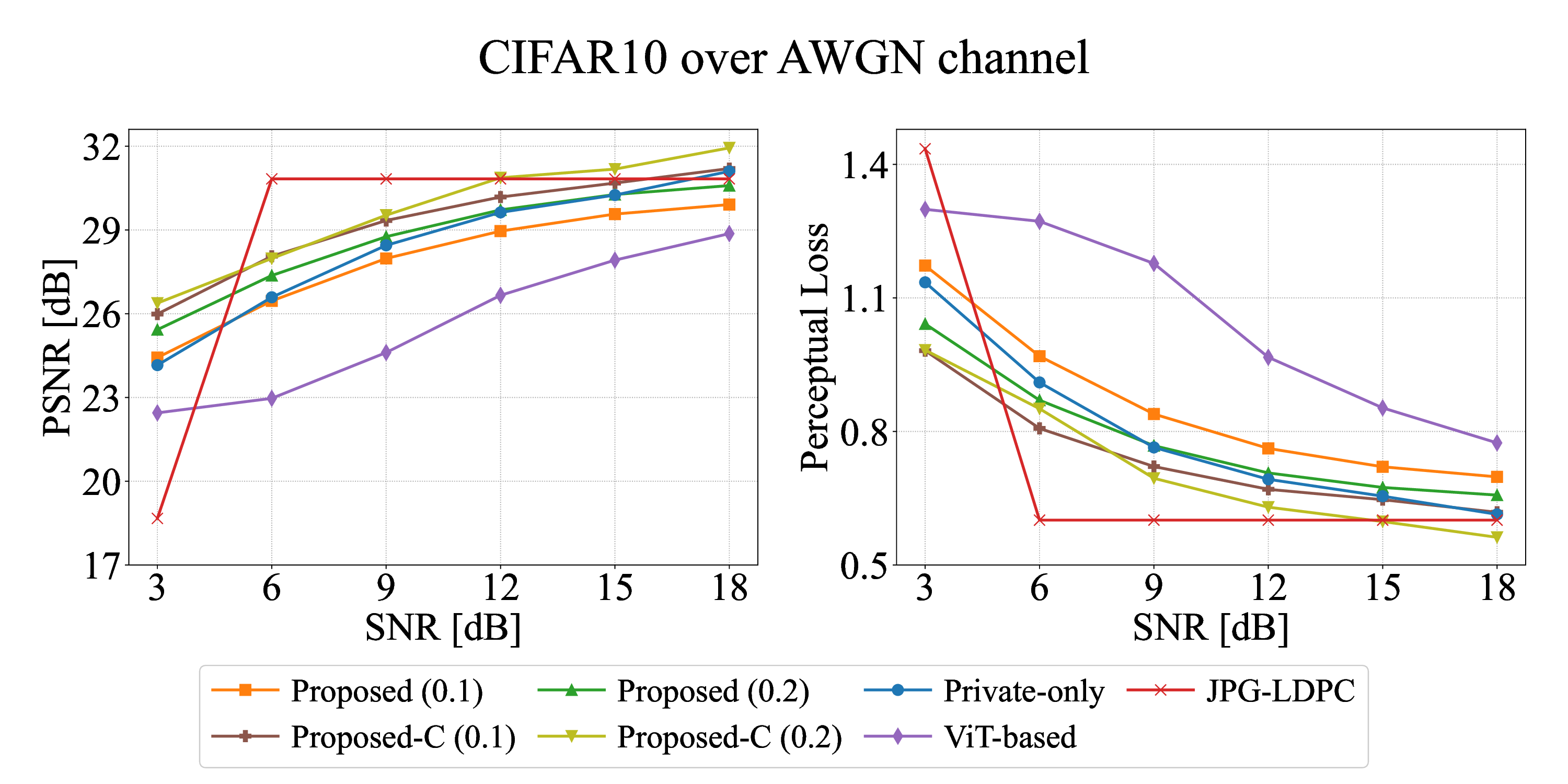}
    \caption{Performance comparison in the AWGN channel: (Left) PSNR, (Right) Perceptual Loss.}
    \label{fig:awgn_results}
\end{figure}
\begin{figure}[t]
    \centering 
    \includegraphics[draft=false, width=\if 1 \mycmd 0.6 \else 1.0 \fi \columnwidth]{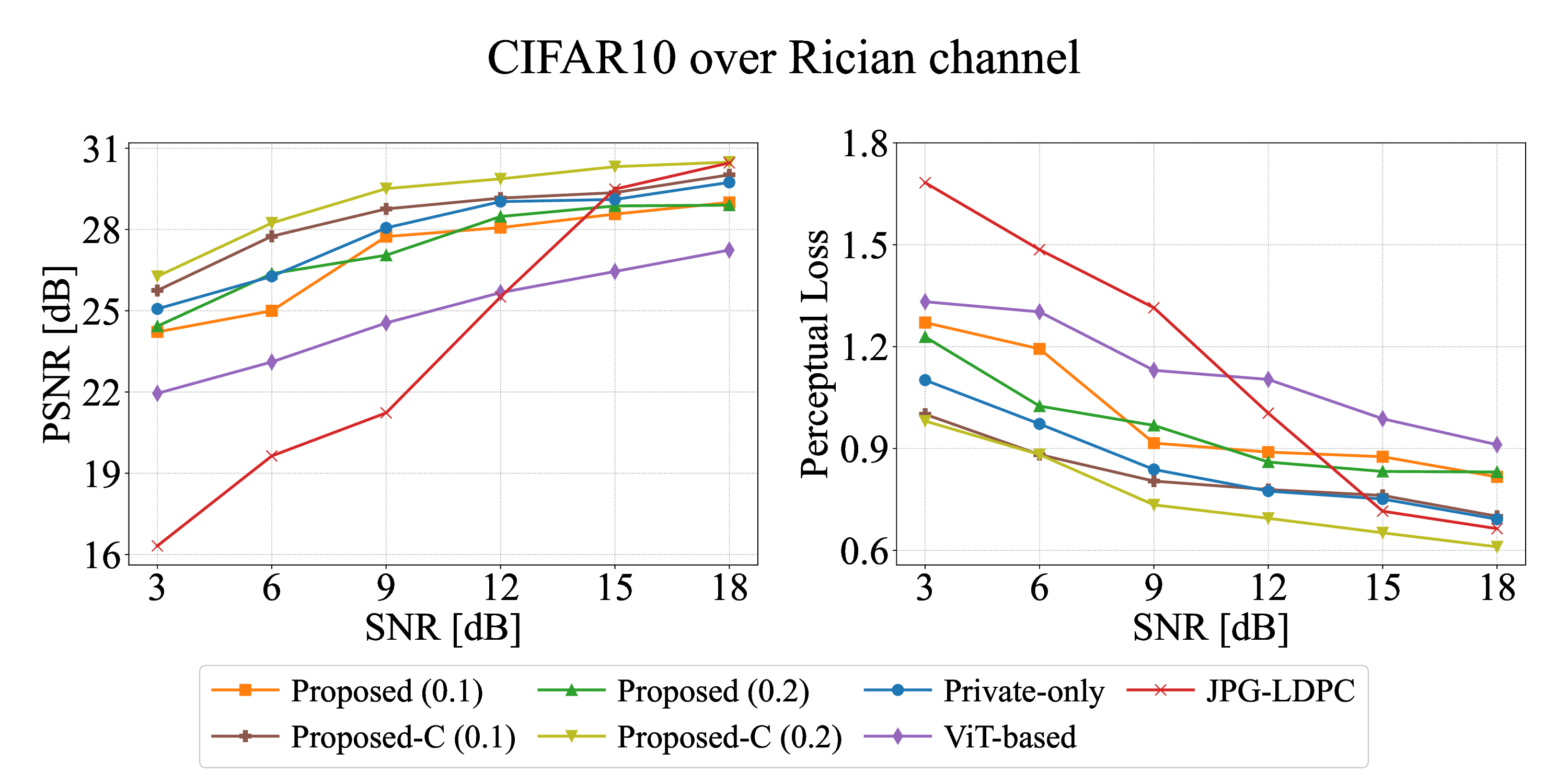}
    \caption{Performance comparison in the Rician channel with \(\mu=1\): (Left) PSNR, (Right) Perceptual Loss.}
    \label{fig:rician_results}
\end{figure}
\begin{figure}[t]
    \centering 
     \includegraphics[draft=false, width=\if 1 \mycmd 0.6 \else 1.0 \fi \columnwidth]{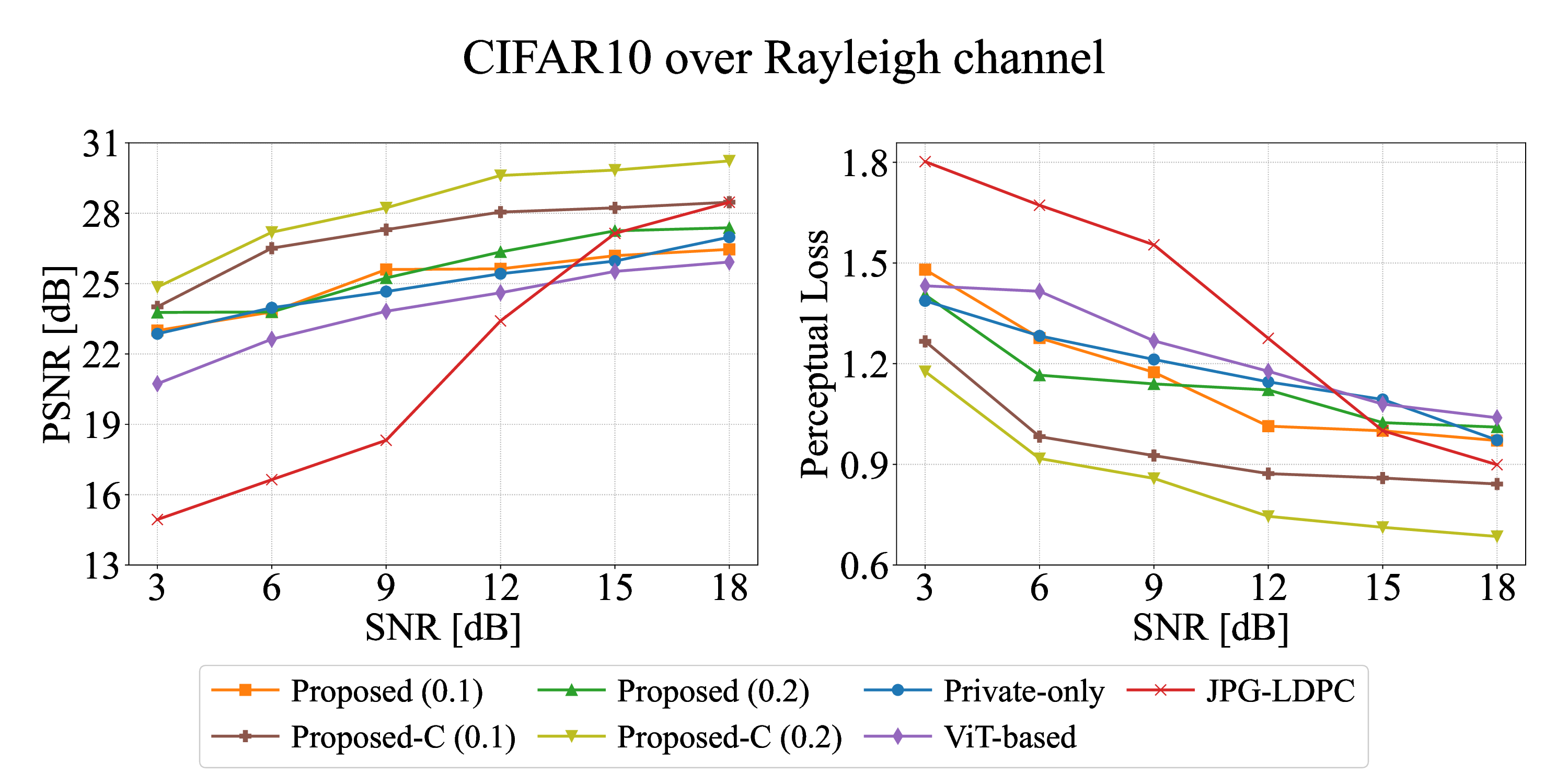}
    \caption{Performance comparison in the Rayleligh channel: (Left) PSNR, (Right) Perceptual Loss.}
    \label{fig:rayleigh_results}
\end{figure}

Fig.~\ref{fig:awgn_results} presents the PSNR and perceptual loss of the CIFAR10 dataset under the AWGN channel using 1024 transmitted symbols per image. The lines marked with ``-C'' indicate the use of the balanced clustering strategy, and the value in parentheses denotes the common feature ratio. Among them, \textbf{Proposed-C (0.2)} achieves the best performance in medium-to-high SNRs (\(\geq 12\) dB), clearly demonstrating the effectiveness of both the proposed private-common feature separation and the balanced clustering strategy. While the LDPC baseline shows better PSNR at moderate SNRs, its performance degrades sharply under low SNRs. In contrast, the proposed semantic multiplexing scheme shows smoother degradation and more stable perceptual quality across all SNR regimes, highlighting its robustness against poor channel qualities. 
Fig.~\ref{fig:rician_results} presents the results under Rician fading, where the proposed methods consistently outperform existing baselines. Notably, \textbf{Proposed-C (0.2)} achieves the highest PSNR across all SNR levels Likewise, Fig.~\ref{fig:rayleigh_results} confirms the superiority of \textbf{Proposed-C (0.2)} under Rayleigh fading, with an even larger performance margin.

Across all three channel conditions, using a higher shared-feature ratio and balanced clustering consistently yields the best reconstruction quality, and the performance gap increases as SNR grows. This suggests that reliable channel conditions favor a larger portion of common features, as they effectively expand symbol resources available for delivering global semantic information.
Conversely, in low-SNR regimes, a smaller shared ratio achieves comparable performance, indicating that allocating more private features helps preserve user-specific semantics under severe channel distortion. 
Nevertheless, in all cases, incorporating group-wise common features consistently outperforms the private-only baseline, demonstrating that even minimal global semantic alignment remains beneficial. Furthermore, performance gap between models with and without clustering confirms that the proposed balanced clustering strategy effectively extracts disentangled common representations across user groups. 





\section{Conclusion}
In this letter, we introduce a multi-user semantic communication framework that extracts private and group-level common features to enable efficient semantic-aware downlink transmission. The framework incorporates a balanced clustering strategy along with a composite loss function that jointly improves semantic separability and reconstruction fidelity by enforcing both feature diversity and accurate recovery. Experimental results across multiple channel environments demonstrate that the proposed approach consistently outperforms existing baselines, achieving the highest reconstruction quality and validating its effectiveness for next-generation semantic communication systems.

\bibliographystyle{IEEEtran}
\bibliography{{IEEEabrv,bibtex}}

\end{document}